\documentclass[3p,times]{elsarticle}

\usepackage{ecrc}

\volume{00}

\firstpage{1}

\journalname{..}

\runauth{}

\usepackage{amssymb}
\usepackage{amsthm}

\usepackage[figuresright]{rotating}

\begin{document}

\begin{frontmatter}



\dochead{}

\title{Zeroing in on jet quenching: a PHENIX perspective}


\author{Jiangyong Jia for the PHENIX Collaboration}

\address{Chemistry Department, Stony Brook University, Stony Brook, NY 11794, USA}
\address{Physics Department, Brookhaven National Laboratory, Upton, NY 11796, USA}

\begin{abstract}
We review recent progresses on jet quenching measurements by
PHENIX. With increased statistics, PHENIX has gone beyond the
single hadron suppression $R_{\rm AA}$, and made measurements on
multiple jet quenching observables, such as $v_2$, $I_{\rm AA}$ and
$v_2^{I_{\rm AA}}$. We argue that, by combining these observables
together, one can achieve a better understanding of the energy loss
mechanism. We present new $\gamma$-hadron correlation results with
associated hadrons extended to low $p_T$; an enhancement has been
observed, suggesting a contribution of genuine medium response that
is relatively unbiased by the initial geometry fluctuations. The
status of full jet reconstruction and future perspective of PHENIX
jet quenching program are discussed.
\end{abstract}

\begin{keyword}
Quark gluon plasma \sep Jet quenching \sep single hadron
suppression \sep di-hadron correlation \sep gamma-hadron
correlation


\end{keyword}

\end{frontmatter}


\section{Introduction}
Jet quenching or suppression of high $p_T$ jets and di-jets is one
of the most intensely studied probes of the strongly-interaction
quark gluon plasma (sQGP) created in Au+Au collisions at RHIC.
Since the suppression level is directly caused by the interaction
of the jets with the medium, jet quenching measurements allow us to
infer the properties of the sQGP. After nearly a decade long
effort, jet quenching, as an experimental phenomena, has been
firmly established at RHIC. However, the exact mechanism for jet
quenching is still under intense debate. The often-used pertubative
QCD (pQCD) framework, which works well in elementary p+p
collisions, fail to describe simultaneously the light and heavy
quark suppression. In contrast, non-perturbative approaches, for
example those based on AdS/CFT gauge gravity duality~\cite{ads},
seem to work well. Even within the pQCD framework, there are many
models that are based on different and often uncontrolled
approximations~\cite{Horowitz:2009eb}, which, when tuned to the
same data, predict very different medium properties.

One way to improve the situation is to study multiple jet quenching
observables at once, capitalizing on their different sensitivities
to the energy loss mechanism. These observables include single
hadron suppression $R_{\rm AA}$ and its azimuthal anisotropy
relative to the reaction plane (RP)  $R_{\rm AA}(\phi-\Psi_{\rm
RP})$ or $v_2$, di-hadron suppression $I_{\rm AA}$ and its
azimuthal anisotropy relative to the RP $I_{\rm AA}(\phi-\Psi_{\rm
RP})$ or $v_2^{I_{\rm AA}}$, as well as modifications of
electron-hadron correlation, $\gamma$-hadron correlations and fully
reconstructed jets. The idea is to fix the energy loss mechanism,
while dramatically varying the path length that the probes
transverse, for example by comparing $R_{\rm AA}$, $v_2$, $I_{\rm
AA}$, and $v_2^{I_{\rm AA}}$ for light hadrons; or fix the path
length and vary the interaction between jet and medium, for example
by comparing di-hadron and electron-hadron correlations. The goal
of this proceedings is to summarize recent PHENIX results on all
these observables, and discuss the physics insights obtained by
combining these results. Reader should refer
to~\cite{Horowitz:2010yi} for a theoretical review on this subject.

\section{Leading hadron suppression: does $R_{\rm AA}$ rise with $p_T$?}
Single hadron suppression $R_{\rm AA}$ is the most studied
observable for jet quenching. Due to the steeply falling spectrum,
this observable suffers from energy loss bias (observed hadron
tends to have small energy loss), which limits its ability in
distinguishing various model scenarios. Nevertheless, as the
$R_{\rm AA}$ measurements become more precise, one can regain some
discriminating power. For example, recent precision measurements of
$\pi^0$ and $\eta$ mesons have started to pin down the shape of the
$R_{\rm AA}$ at high $p_T$. This is important for understanding the
energy loss mechanism. A slow rise of the $R_{\rm AA}$, for
instance, is expected for radiative energy loss due to its
logarithmic dependence on the initial jet energy.

\begin{figure}[h]
\begin{tabular}{lr}
\begin{minipage}{0.80\linewidth}
\begin{flushleft}
\includegraphics[width=0.33\textwidth,angle=90]{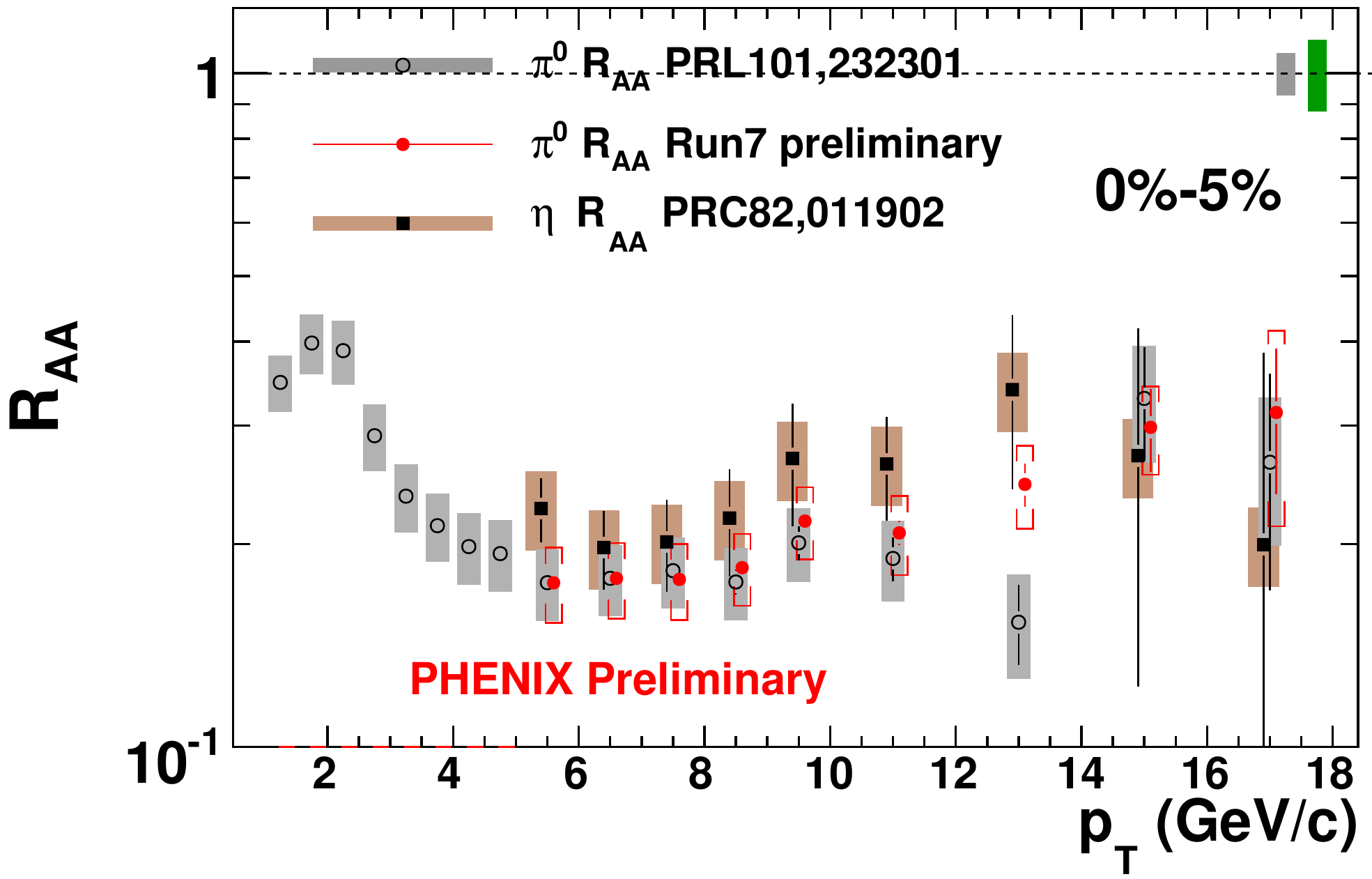}\includegraphics[width=0.33\textwidth,angle=90]{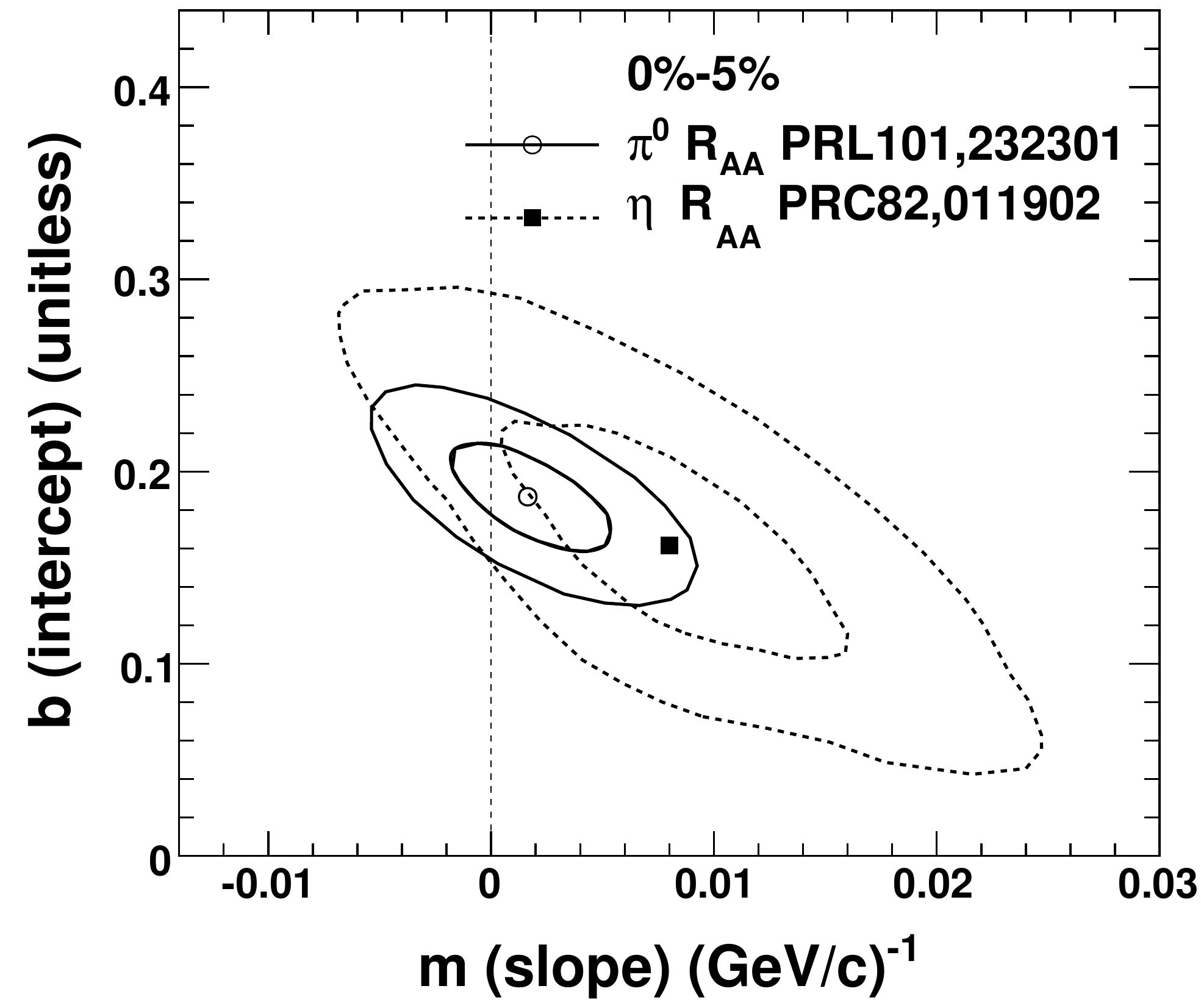}
\end{flushleft}
\end{minipage}
\hspace*{-0.2in}\begin{minipage}{0.20\linewidth}
\begin{flushright}
\caption{\label{fig:spectra} (Left) $R_{\rm AA}$ for $\pi^0$ and $\eta$ mesons from different runs in 0-5\% centrality bin;
(Right) 1 and 2 standard deviation contours of the intercept vs slope for published RUN4 $\pi^0$ and RUN7 $\eta$~\cite{Adare:2008qa}
(The contour parameters for RUN7 $\eta$ $R_{\rm AA}$ were converted to the function used for RUN4 $\pi^0$ via a simple
linear transformation).}
\end{flushright}
\end{minipage}
\end{tabular}
\end{figure}

Results for the 0-5\% centrality bin are shown in
Fig.~\ref{fig:spectra}, where RUN4 $\pi^0$ and RUN7 $\eta$ $R_{\rm
AA}$~\cite{Adare:2008qa} are fitted with a linear function $R_{\rm
AA}= b\;+\;m\;p_T$ to extract the slope parameter $m$. The data do
indicate a small positive slope, but the significance is less than
$1\sigma$ for RUN4 $\pi^0$ and slightly above $1\sigma$ for $\eta$.
The RUN7 $\pi^0$ $R_{\rm AA}$ seem to suggest somewhat more
significant increasing trend with $p_T$. The updated numbers should
come out soon.

\section{Azimuthal anisotropy of leading hadron suppression: what is the
path length dependence of energy loss?} The overlap zone of the
Au+Au collision is not symmetric; consequentially, the emission
rate of high $p_T$ particle may vary with its angle relative to the
reaction plane (RP) $\phi-\Psi_{\rm RP}$. This azimuthal anisotropy
can be characterized by either $v_2$ parameter or
$R_{AA}(\phi-\Psi_{\rm RP})$, whose magnitudes directly reflect the
path length dependence of the energy loss.

\begin{figure}[h]
\begin{tabular}{lr}
\begin{minipage}{0.80\linewidth}
\begin{flushleft}
\includegraphics[width=0.4\textwidth,angle=90]{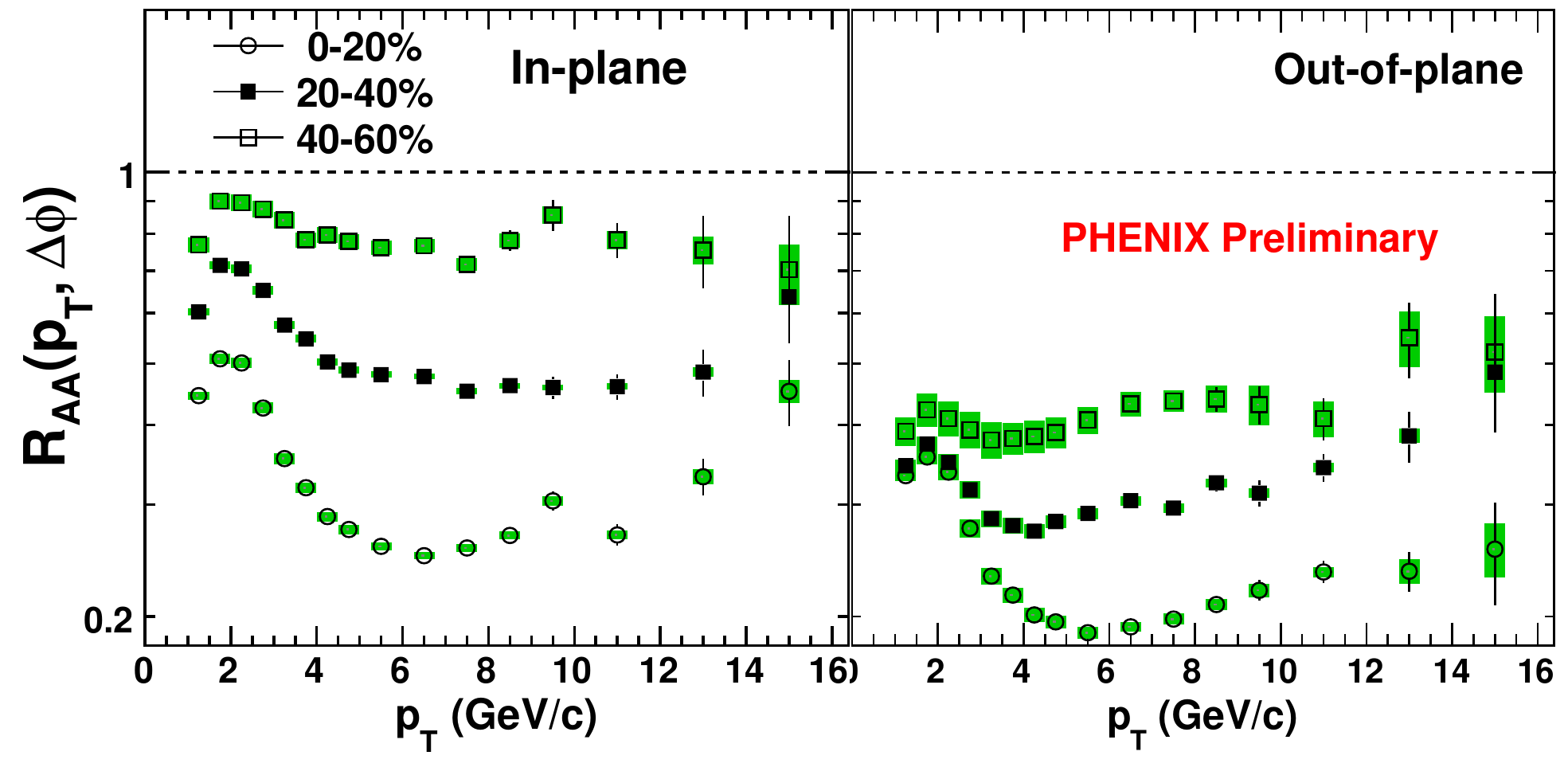}\end{flushleft}
\end{minipage}
\hspace*{-0.2in}\begin{minipage}{0.20\linewidth}
\begin{flushright}
\caption{\label{fig:raaphi} $R_{\rm AA}$ vs $p_T$ in the in-plane (left) and out-of-plane (right) directions, results for three centrality bins are shown.}
\end{flushright}
\end{minipage}
\end{tabular}
\end{figure}

Figure~\ref{fig:raaphi} shows the $p_T$ dependence of the $R_{\rm
AA}$ separately for in-plane ($R_{\rm AA}^{\rm in}$) and
out-of-plane ($R_{\rm AA}^{\rm out}$). At low $p_T$, the $R_{\rm
AA}^{in}$ varies by almost a factor of 2 while that for
out-of-plane direction is almost unchanged; this can be interpreted
as a stronger radial flow influence in the in-plane direction. At
high $p_T$, the $R_{\rm AA}^{\rm out}$ is more suppressed than
$R_{\rm AA}^{\rm in}$, reflecting greater path length for
out-of-plane going jets. Interestingly, $R_{\rm AA}^{\rm out}$ also
shows a stronger $p_T$ dependence than the in-plane direction,
suggesting that the $R_{\rm AA}^{\rm out}$ can better expose the
true shape of the $R_{\rm AA}$ at high $p_T$ from jet quenching.

\begin{figure}[h]
\begin{tabular}{lr}
\begin{minipage}{0.80\linewidth}
\begin{flushleft}
\centerline{\includegraphics[width=0.7\textwidth]{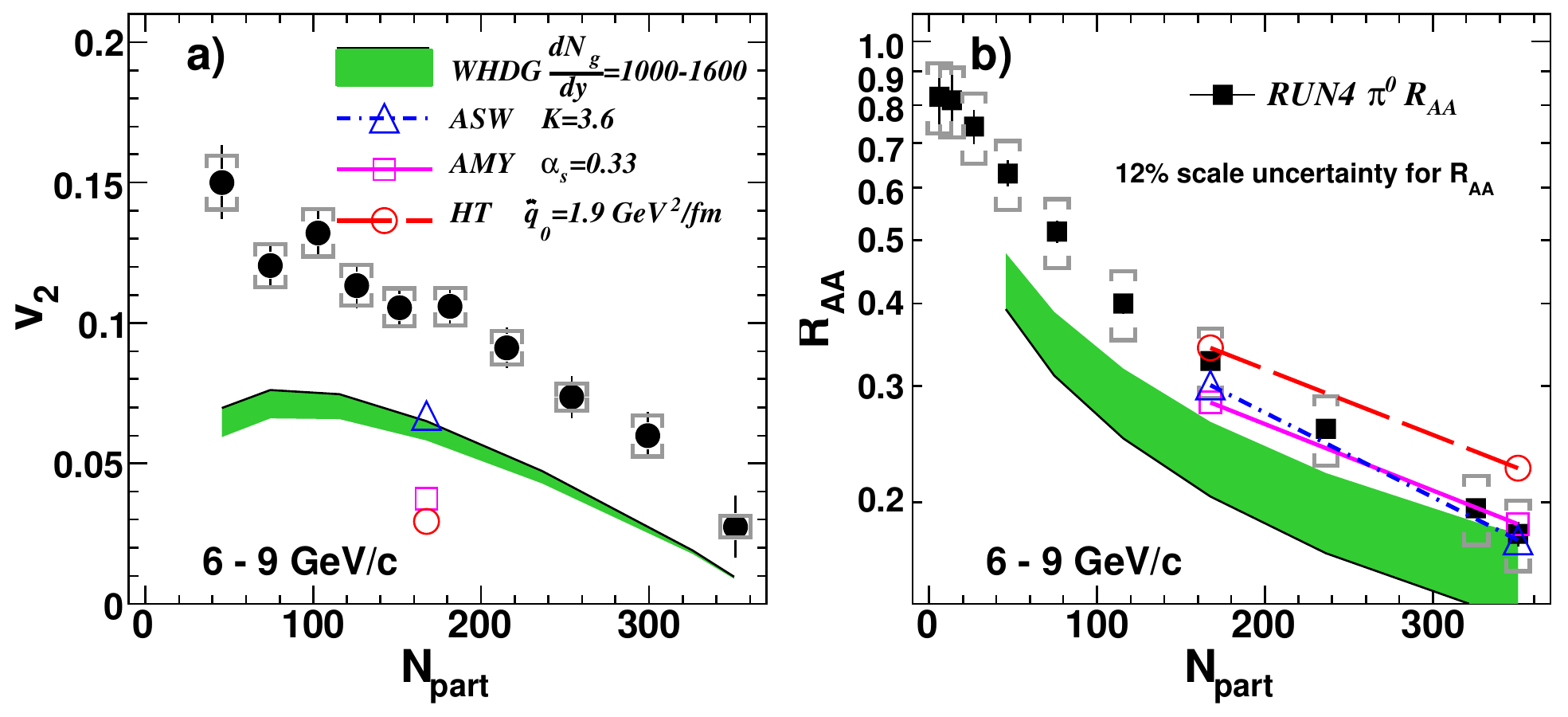}}
\vspace*{-0.285in}\centerline{\includegraphics[width=0.7\textwidth]{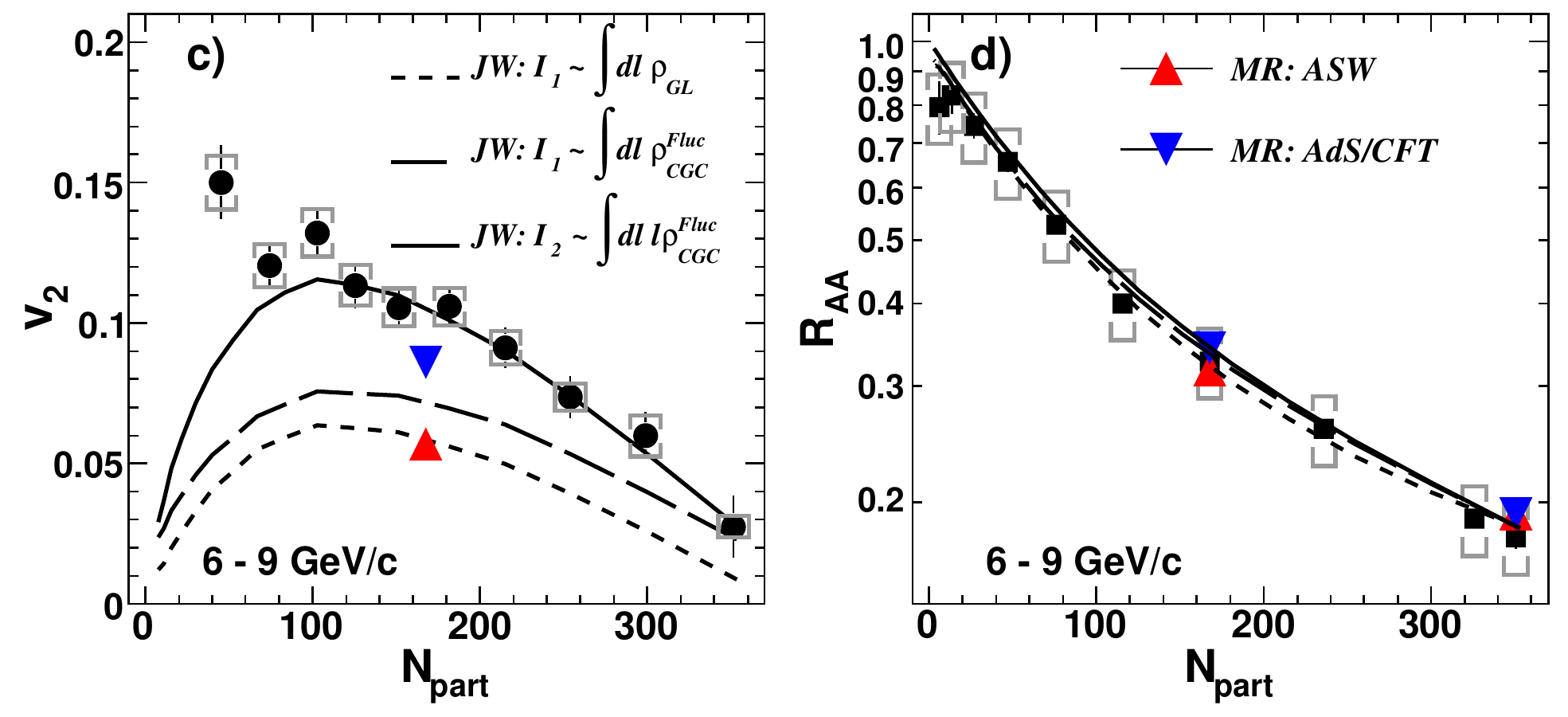}}
\end{flushleft}
\end{minipage}
\hspace*{-0.2in}\begin{minipage}{0.20\linewidth}
\begin{flushright}
\caption{\label{fig:v2a} The centrality dependence of $v_2$ (left panels) and $R_{\rm AA}$ (right panels) in 6-9 GeV/$c$ range. They are compared with various
pQCD models calculations (top panels), and schematic calculations that explore the roles of geometry and cubic path length dependence (right panels). }
\end{flushright}
\end{minipage}
\end{tabular}
\end{figure}

Figure~\ref{fig:v2a}a shows the centrality dependence of $v_2$ at
high $p_T$~\cite{Adare:2010sp}. The $v_2$ values are large and
increase toward peripheral collisions. They are compared to various
pQCD model calculations~\cite{Bass:2008rv}. These models are
calculated with different geometry and different implementation of
energy loss processes. For example, the HT and ASW models include
only coherent radiative energy loss, while the AMY and WHDG models
also include collisional energy loss. The calculations are tuned to
the $R_{\rm AA}$ value in the 0-5\% centrality bin (right panel),
thus the centrality dependence of $v_2$ and $R_{\rm AA}$ are
predictions. All models describe the centrality dependence of
$R_{\rm AA}$ reasonably well, but significantly under-predict the
$v_2$ data. Furthermore, the calculated $v_2$ values differ among
themselves. Since $R_{\rm AA}$ and $v_2$ are anti-correlated, ${\it
i.e.}$~a small $R_{\rm AA}$ implies a large $v_2$ and vice versa,
it is unlikely that one can describe both $v_2$ and $R_{\rm AA}$ by
simply re-tuning the quenching parameters in these models.

How to resolve this apparent discrepancies between data and
theories?  A recent estimation~\cite{Jia:2010ee} shows that the
calculated $v_2$ could be increased by 30-40\% by considering
modifications of collision geometry due to event-by-event
fluctuations and CGC effects (dashed line in Fig.~\ref{fig:v2a}c).
However, the calculation still falls below the data. Unless there
are other modifications of collision geometry that we are not aware
of, this large discrepancy implies that our current picture of
pQCD-based energy loss is not complete. For example, the quadratic
$l$ dependence formula for radiative energy loss, $\Delta
E\propto\int_{\tau_0}^{\infty}d\tau\frac{\tau-\tau_0}{\tau_0}\rho(\tau,\mathbf{r}+\mathbf{n}\tau)
\propto\int_{0}^{\infty}d\tau\rho_{\rm
part}(\mathbf{r}+\mathbf{n}\tau)\sim l^2$ (where $\rho_{\rm part}$
is participant density), may need to be changed to increase the
anisotropy. There are three ideas along this line of reasoning.
Liao and Shuryak~\cite{Liao:2008dk} do this by requiring that most
of the energy loss in sQGP be concentrated around $T_c$. One can
increase the $v_2$ by increasing the formation time $\tau_0$ to
$1.5-2.0$ fm/$c$~\cite{Pantuev:2005jt}. One can also increase the
$v_2$ with a cubic $l$ dependence of energy loss: $\Delta
E\propto\int_{\tau_0}^{\infty}d\tau\left(\frac{\tau-\tau_0}{\tau_0}\right)^2\rho(\tau,\mathbf{r}+\mathbf{n}\tau)
\propto\int_{0}^{\infty}d\tau\tau\rho_{\rm
part}(\mathbf{r}+\mathbf{n}\tau) \sim l^3$; such dependence is
expected in certain non-perturbative energy loss calculations based
on AdS/CFT gravity-gauge dual theory~\cite{Dominguez:2008vd}.
Indeed, Fig.~\ref{fig:v2a}c shows that our data agree well with
calculations~\cite{Marquet:2009eq,Jia:2010ee} based on such $l$
dependence.

\begin{figure}[h]
\begin{tabular}{lr}
\begin{minipage}{0.80\linewidth}
\begin{flushleft}
\centerline{\includegraphics[width=0.35\textwidth,angle=90]{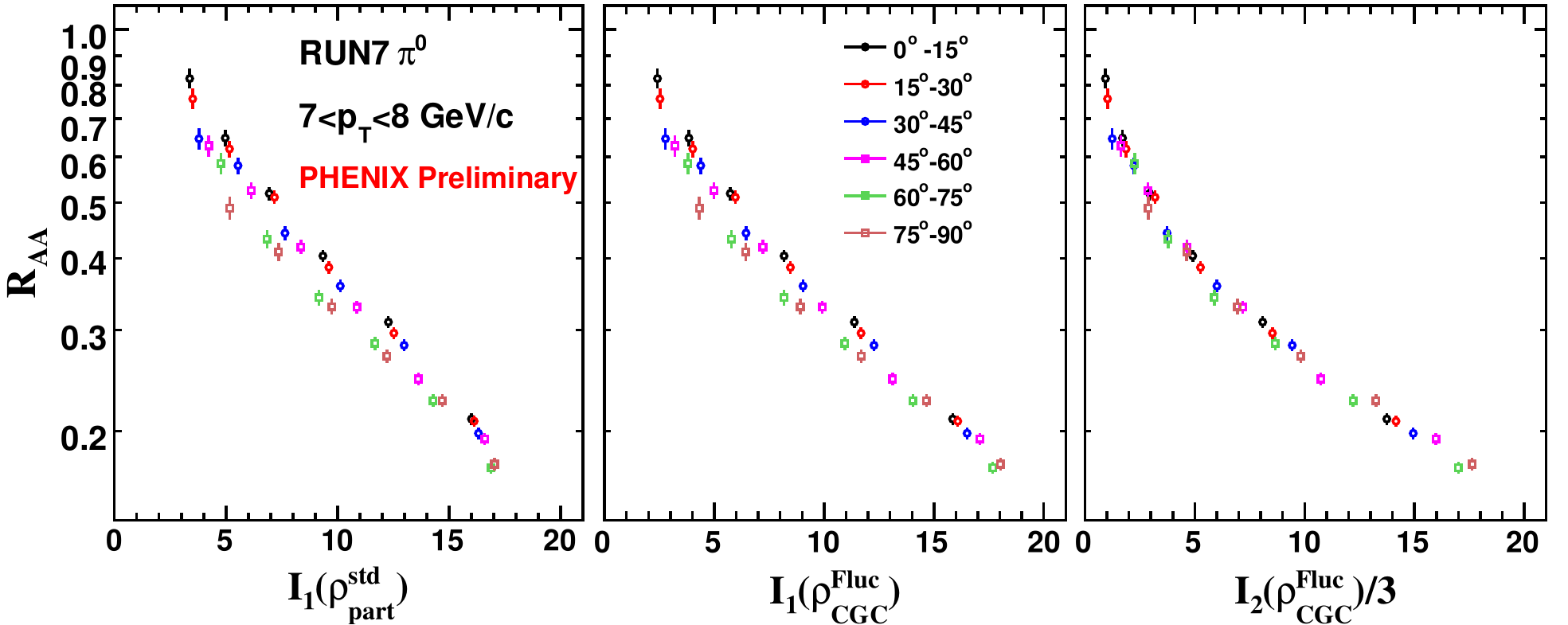}}
\end{flushleft}
\end{minipage}
\hspace*{-0.2in}\begin{minipage}{0.20\linewidth}
\begin{flushright}
\caption{\label{fig:v2b}  The $R_{\rm AA}$ vs $I_{m}=
\int_{0}^{\infty}d\tau \tau^{m-1}\rho_{\rm
part}(\mathbf{r}+\mathbf{n}\tau)$ ($m=1,2$) for 6 centrality and 6 angle bins, calculated for standard Glauber geometry (left), CGC geometry with E-by-E fluctuation (middle and right).}
\end{flushright}
\end{minipage}
\end{tabular}
\end{figure}

To further pinpoint the influence of the collision geometry and the
path length dependence to high $p_T$ anisotropy, we studied the $l$
scaling behavior of the $R_{\rm AA}$ for various centrality and
$\phi-\Psi_{\rm RP}$ angle bins. The idea is that if two selections
have different $N_{\rm part}$ and $\phi-\Psi_{\rm RP}$, but similar
average energy loss $\langle\Delta E\rangle$, their suppression
levels should be similar. Thus if real energy loss scales as
$\Delta E\propto \int_{0}^{\infty}d\tau \tau^{m-1}\rho_{\rm
part}(\mathbf{r}+\mathbf{n}\tau)\equiv I_{m}$ ($m=1,2$), then one
expects $R_{\rm AA}\propto\langle I_{m}\rangle$. The result of this
exercise is shown in Fig.~\ref{fig:v2b} for 6 centrality and 6
angles bins, that is $R_{\rm AA}$ are plotted versus $I_{m}$ at
high $p_T$ (7-8 GeV/$c$). $I_1(\rho_{\rm part}^{\rm std})$
corresponds to quadratic $l$ dependence in standard Glauber
geometry, $I_1(\rho_{\rm CGC}^{\rm Fluc})$ corresponds to quadratic
$l$ dependence in CGC geometry with E-by-E fluctuations, and
$I_2(\rho_{\rm CGC}^{\rm Fluc})$ corresponds to cubic $l$
dependence in CGC geometry with E-by-E fluctuations. Comparing the
three panels, we see that the $R_{\rm AA}$ does not scale with
$I_1(\rho_{\rm part}^{\rm std})$ and $I_1(\rho_{\rm CGC}^{\rm
Fluc})$, although the latter is slightly better; but it scales very
well with $I_2(\rho_{\rm CGC}^{\rm Fluc})$.

\begin{figure}[h]
\begin{tabular}{lr}
\begin{minipage}{0.80\linewidth}
\begin{flushleft}
\centerline{\includegraphics[width=0.94\textwidth]{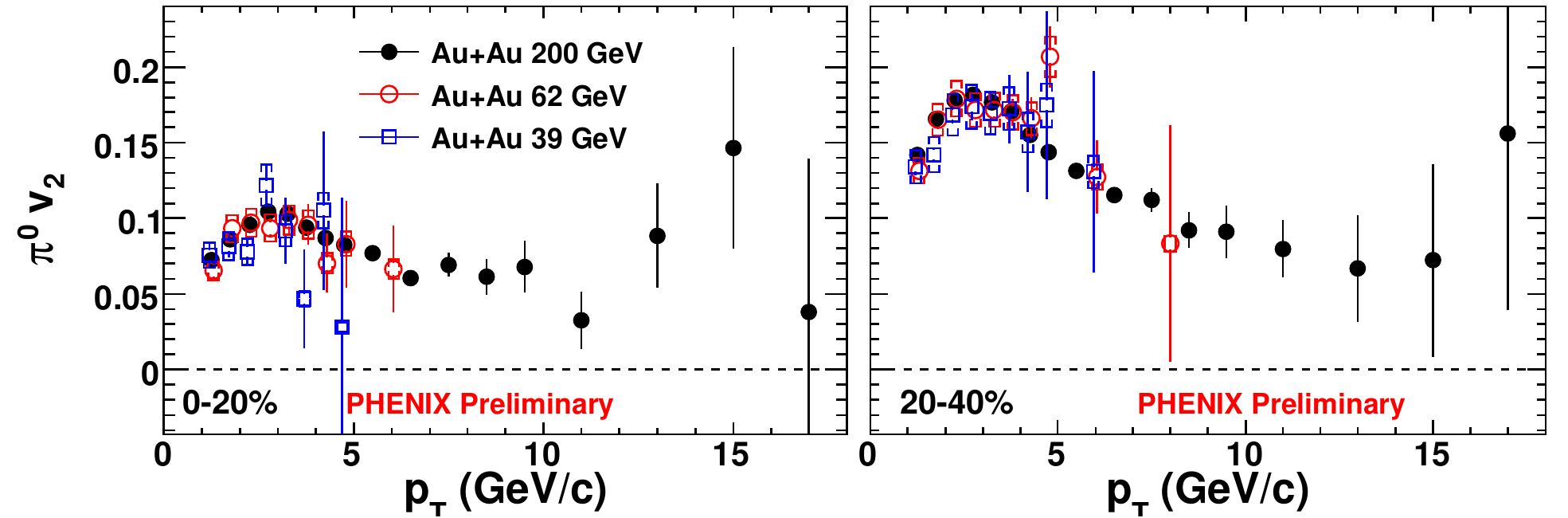}}
\end{flushleft}
\end{minipage}
\hspace*{-0.17in}\begin{minipage}{0.17\linewidth}
\begin{flushright}
\caption{\label{fig:v2c} The $\pi^0$ $v_2$ versus $p_T$ for $\sqrt{s_{NN}}=200,62 $ and 39 GeV. }
\end{flushright}
\end{minipage}
\end{tabular}
\end{figure}

Figure~\ref{fig:v2c} shows the energy dependence of $v_2$ from
$\sqrt{s_{\rm NN}}=39-200$ GeV. The $v_2$ at $\sqrt{s_{\rm
NN}}=200$ GeV shows a gradual drop from 3 GeV/$c$ to 7-10 GeV/$c$
and remain positive at higher $p_T$. The $v_2$s at $\sqrt{s_{\rm
NN}}=39$ and 62 GeV have limited $p_T$ reach, but their magnitudes
are consistent with 200 GeV results, both at low $p_T<2$ GeV/$c$
where collective flow dominates and at $p_T>4$ GeV/$c$ where jet
quenching should play a significant role. The former result
suggests that, even at low energies, the medium already has to
thermalize quickly and have small dissipation. The latter result is
quite surprising, since we know that the $R_{\rm AA}$s at low
energies are less suppressed than at $\sqrt{s_{\rm NN}}=200$ GeV.
It would be interesting to see how well the pQCD models, which
describe the $R_{\rm AA}$ at low $\sqrt{s_{\rm
NN}}$~\cite{Adare:2008cx}, can reproduce the $v_2$.

\section{Di-hadron suppression $I_{AA}$ and its azimuthal anisotropy $I_{AA}(\phi-\Psi_{RP})$: Further
probing into the path length dependence of energy loss with
away-side jet}

\begin{figure}[h]
\begin{tabular}{lr}
\begin{minipage}{0.8\linewidth}
\begin{flushleft}
\centerline{\includegraphics[width=0.32\textwidth,angle=90]{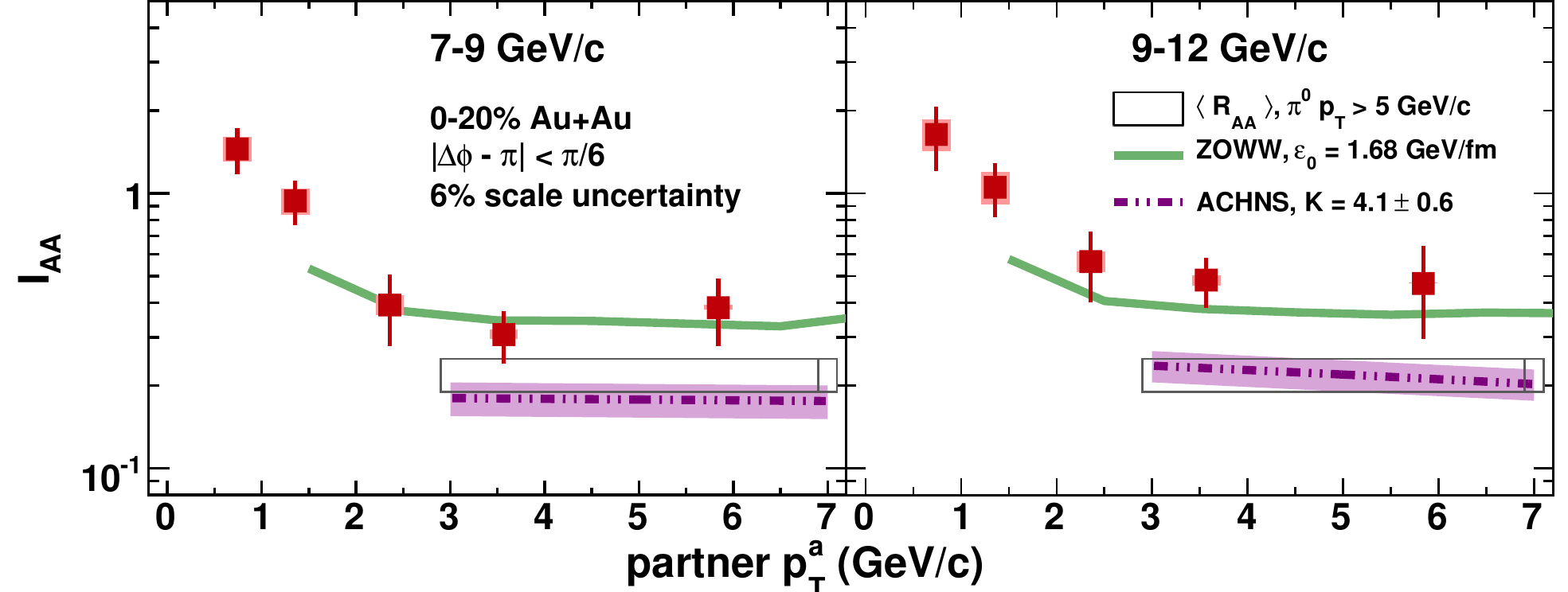}}
\end{flushleft}
\end{minipage}
\hspace*{-0.20in}
\begin{minipage}{0.2\linewidth}
\begin{flushright}
\caption{\label{fig:iaaa} Away-side per-trigger yield suppression $I_{\rm AA}$ as function of associated hadron $p_T$ in central Au+Au collisions.}
\end{flushright}
\end{minipage}
\end{tabular}
\end{figure}

High $p_T$ hard-scattered jets, at leading order, are produced in
pairs that are back-to-back in the azimuthal direction. One
observable of interest is the suppression pattern of the away-side
jet opposite to the trigger hadron above a certain $p_T$ threshold.
This suppression is quantified by $I_{\rm AA}$, the ratio of the
per-trigger yield (away-side jet multiplicity normalized by number
of triggers) in Au+Au collisions to that in p+p collisions. Pure
geometrical consideration would suggest $I_{\rm AA}<R_{\rm AA}$ due
to a longer path length traversed by the away-side jet. But a
recent PHENIX measurement~\cite{Adare:2010ry} shows that $I_{\rm
AA}$ is constant for associated hadron $p_T>3$ GeV/$c$, and this
constant level is above the $R_{\rm AA}$ for the trigger hadrons,
i.e. $I_{\rm AA}>R_{\rm AA}$ (see Fig.~\ref{fig:iaaa}).
Furthermore, the constant level of $I_{\rm AA}$ becomes even less
suppressed for higher trigger $p_T$. This result can be explained,
at least partially, by the bias of the away-side jet energy by the
trigger $p_T$: the initial away-side jet spectra are harder for
higher trigger $p_T$, consequentially, a larger fractional energy
loss is required for the same $I_{\rm AA}$ value. This result rules
out the pure jet attenuation scenario where the jet survival
probability depends only on the path length.

The data in Figure~\ref{fig:iaaa} are compared with several pQCD
model calculations. The quenching parameters of these calculations
have been tuned to reproduce the $R_{\rm AA}$ data, so the
calculations can be regarded as predictions. The ACHNS model (based
on ASW framework) tends to predict $I_{\rm AA}\lesssim R_{\rm AA}$,
thus disagreeing with the data, while ZOWW model (based on HT
framework) can describe the data rather well. This is another
example showing that one can discriminate models by combing
multiple experimental observables.

\begin{figure}[h]
\begin{tabular}{lr}
\begin{minipage}{0.8\linewidth}
\begin{flushleft}
\centerline{\includegraphics[width=0.85\textwidth]{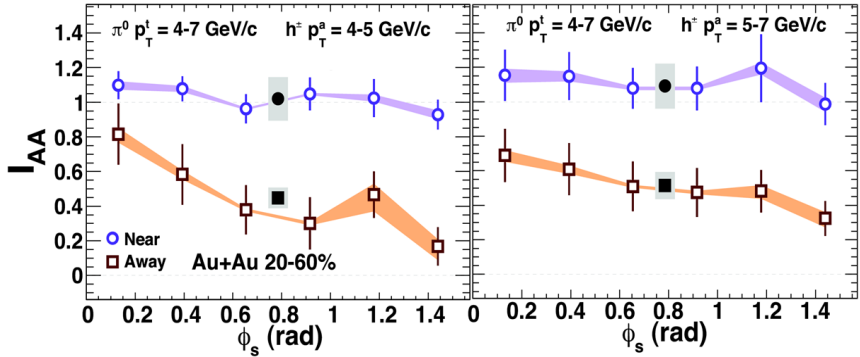}}
\end{flushleft}
\end{minipage}
\hspace*{-0.20in}\begin{minipage}{0.2\linewidth}
\begin{flushright}
\caption
{\label{fig:iaab} $I_{\rm AA}$ as function of angle relative to reaction plane $\phi_s=\phi-\Psi_{RP}$
in two associated hadron $p_T$ bins (left: 4-5 GeV/$c$, right: 5-7 GeV/$c$)
at both the near- (circles) and the away-side (boxes).}
\end{flushright}
\end{minipage}
\end{tabular}
\end{figure}

PHENIX also measured the anisotropy of the away-side suppression,
that is $I_{\rm AA}$ as a function of angle relative to the RP (see
Fig.~\ref{fig:iaab})~\cite{Adare:2010mq}. While the near-side
$I_{\rm AA}$ is essentially unmodified, the away-side $I_{\rm AA}$
shows a strong variation with $\phi_s=\phi-\Psi_{\rm RP}$,
sometimes by more than factor of two from in-plane to out-of-plane.
This variation corresponds to an anisotropy parameter of
$v_2^{I_{\rm AA}}=0.29_{-0.11}^{+0.15}$, and is much bigger than
that for inclusive $\pi^0$ in the same trigger $p_T$ range
$v_2=0.13\pm0.01$. This measurement is statistics limited, however
if the result holds, it would have severe consequence for energy
loss models since they usually predict much smaller anisotropy (for
example, ASW calculation predicts $v_2^{I_{\rm AA}}<0.05$).

Before closing this section, we want to point out that the four
observables, $R_{\rm AA}$, $v_2$, $I_{\rm AA}$ and $v_2^{I_{\rm
AA}}$, are intrinsically correlated. A smaller $R_{\rm AA}$
naturally implies a larger $v_2$, and a smaller $I_{\rm AA}$
implies a larger $v_2^{I_{\rm AA}}$. The relation between $R_{\rm
AA}$ and $I_{\rm AA}$ also depends on the away-side spectra shape
controlled by trigger $p_T$, but in general a smaller $R_{\rm AA}$
implies a smaller $I_{\rm AA}$. Thus it is rather surprising to see
that $I_{\rm AA}$ is less suppressed than $R_{\rm AA}$ ($I_{\rm
AA}>R_{\rm AA}$), yet has a larger anisotropy ($v_2^{I_{\rm
AA}}>v_2$).
\section{Non-photonic electron-hadron correlation: probing the in-medium modifications of heavy quark jets}
The observation of large suppression for non-photonic single
electron (NPE), at a level similar to that for inclusive hadrons,
remains a challenge for pQCD models~\cite{Adare:2006nq}. Since most
of these electrons come from semi-leptonic decay of charm and
bottom mesons, the large suppression suggests that the charm and
bottom quarks interact with the sQGP much more than expected from
pQCD models. Non-photonic electron-hadron correlations provide
valuable complementary information for heavy quark energy loss.
This is because a large fraction of heavy quarks are produced in
back-to-back pair at RHIC energy, and the fragmentation of the
companion heavy quark is expected to contribute significantly to
away-side hadrons associated with triggering electrons.

\begin{figure}[h]
\begin{tabular}{lr}
\begin{minipage}{0.83\linewidth}
\begin{flushleft}
\centerline{\includegraphics[width=0.5\textwidth]{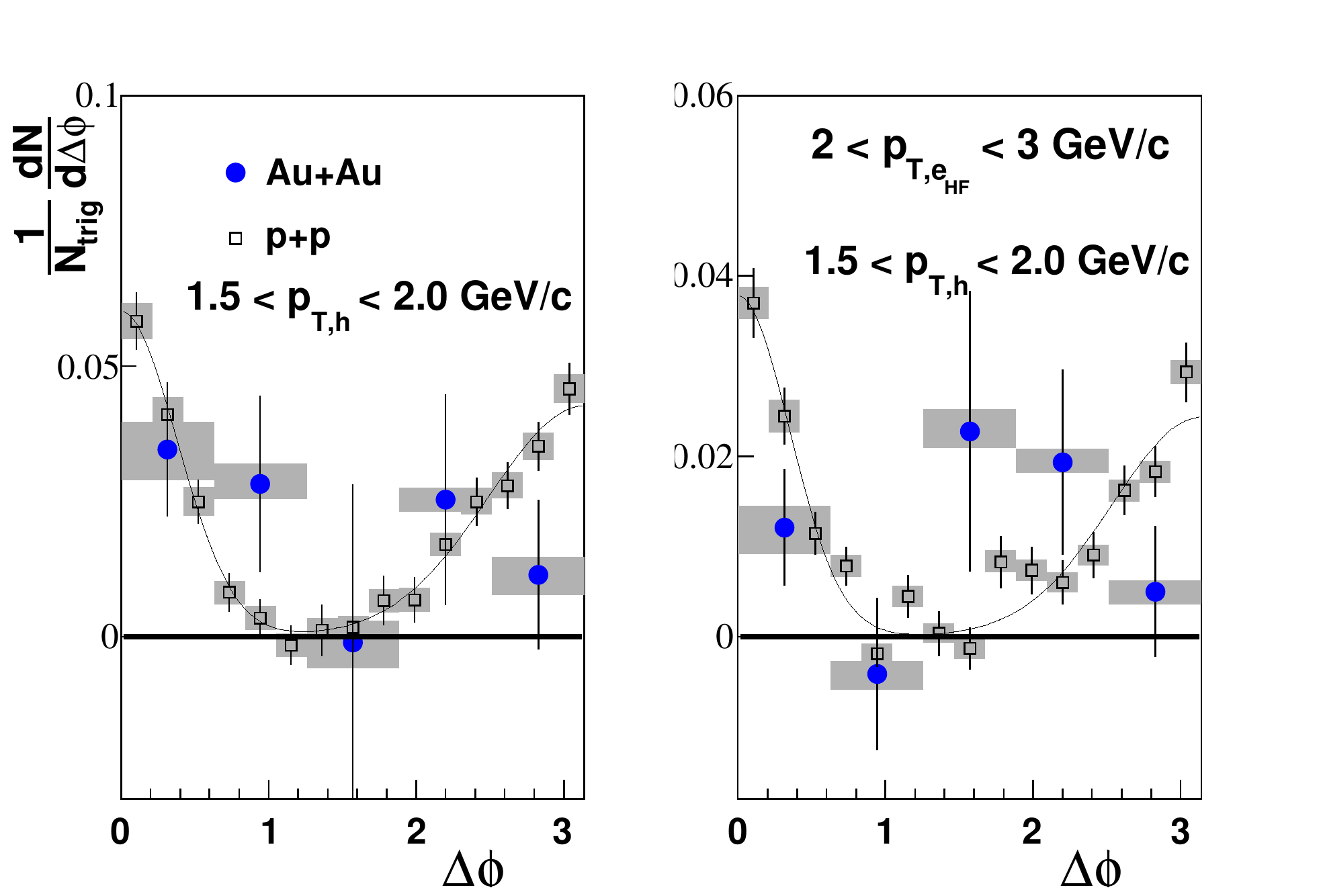}\includegraphics[width=0.5\textwidth]{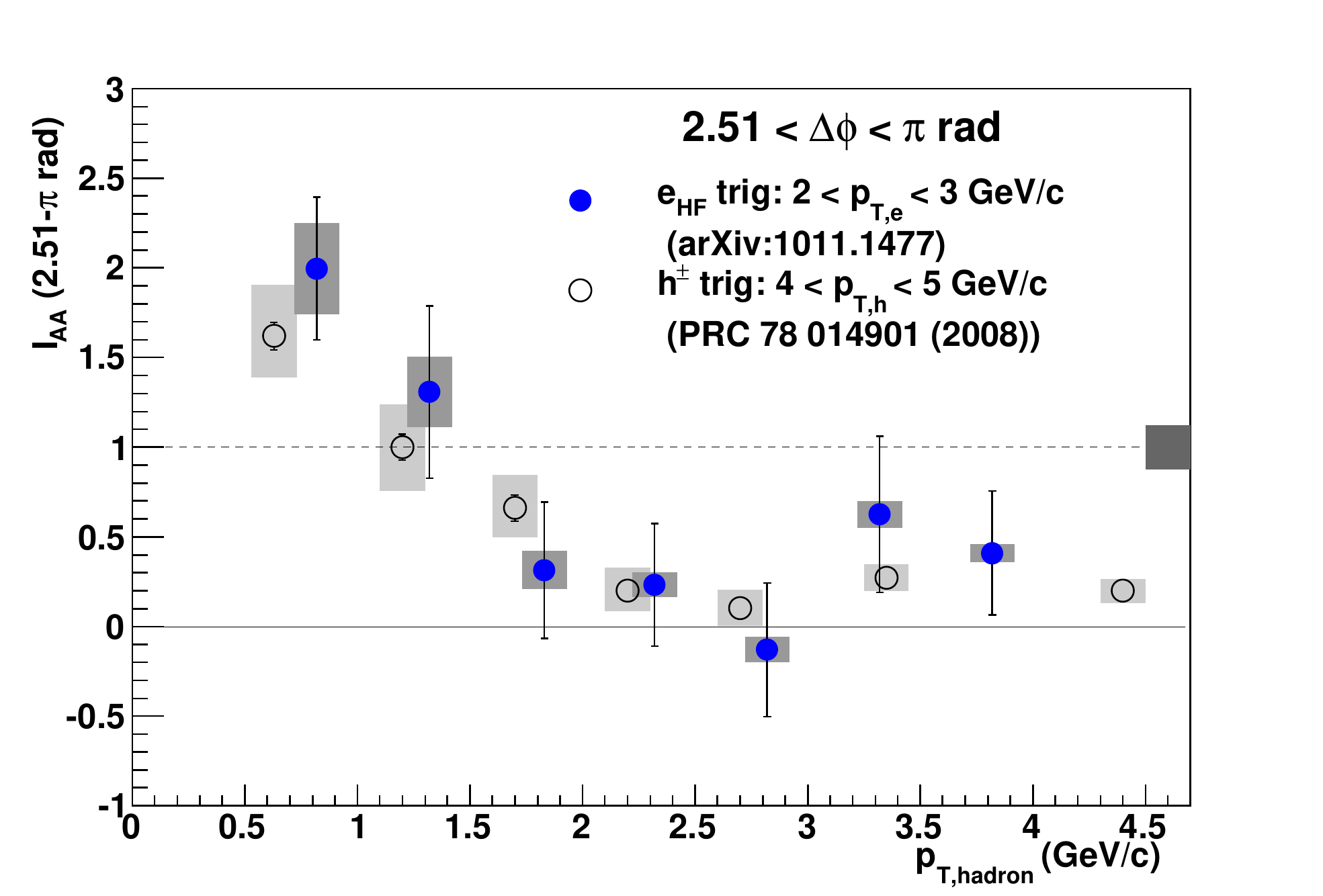}}
\end{flushleft}
\end{minipage}
\hspace*{-0.20in}\begin{minipage}{0.17\linewidth}
\begin{flushright}
\caption{\label{fig:electronh} Non-photonic electron-hadron correlation: (left) Per-trigger yield azimuthal angle distributions for p+p and Au+Au collisions, and (right) $I_{AA}$}
\end{flushright}
\end{minipage}
\end{tabular}
\end{figure}

Results from a first measurement of the NPE-hadron correlation from
PHENIX~\cite{Adare:2010ud} (see Fig.~\ref{fig:electronh}) suggests
that the away-side hadrons are strongly modified relative to p+p.
The away-side hadron yield, when integrated in a $\pm30^o$ window
around $\Delta\phi=\pi$, indicates an enhancement below 1.5 GeV/$c$
and a suppression above 1.5 GeV/$c$. Unfortunately, the large
statistical uncertainties do not allow us to conclude whether the
modification patterns are the same as that for di-hadron
correlations~\cite{Adare:2008cqb}. PHENIX has installed a silicon
vertex detector (VTX) for the next run (RUN11). VTX has the ability
to directly reconstruct and distinguish between charm and bottom
mesons. This should allow us to directly correlate charm and bottom
mesons with charged hadrons. We expect PHENIX to carry out the
first measurement of charm and bottom separated heavy flavor
suppression and correlation measurements in RUN11.

\section{Direct $\gamma$-hadron correlation: an unbiased probe for
jet quenching and medium response} Our discussions of the jet
quenching so far have been focused on the single hadron and
di-hadron correlation observables. These observables are subject to
energy loss bias and geometrical bias, and they represent a
complicated convolution of jets with different initial energy and
different energy loss. In contrast, direct $\gamma$ and fully
reconstructed jet are much less affected by these biases. For
example, one can gauge the away-side jet energy with a direct
$\gamma$ trigger, and one can systematically control the surface
bias by studying the jet $R_{\rm AA}$ as a function of jet cone
size. $\gamma$-hadron and jet-hadron correlations also give us
direct access to jet modification and medium response.

\begin{figure}[h]
\begin{tabular}{lr}
\begin{minipage}{0.75\linewidth}
\begin{flushleft}
\centerline{\includegraphics[width=0.42\textwidth,angle=90]{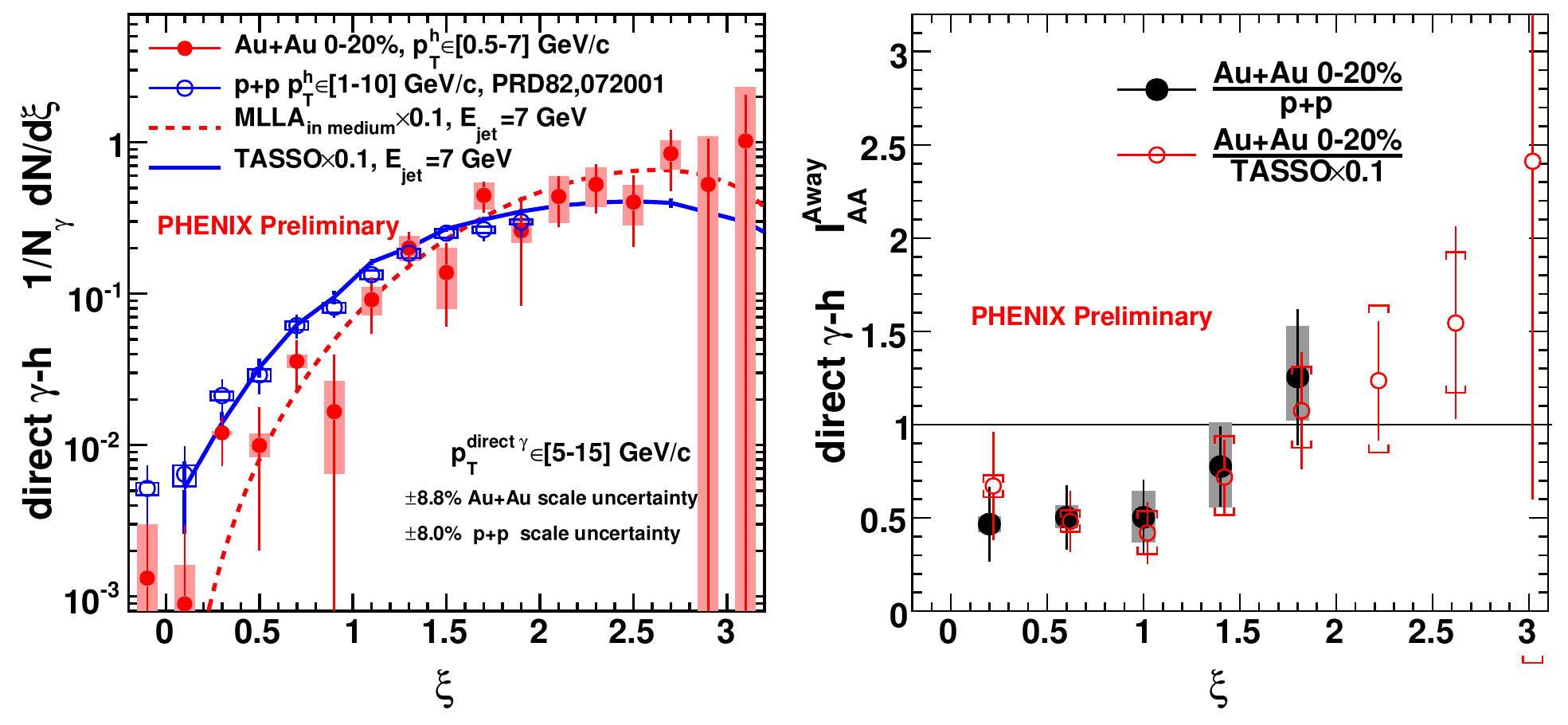}}
\end{flushleft}
\end{minipage}
\hspace*{-0.10in}\begin{minipage}{0.22\linewidth}
\begin{flushright}
\caption{\label{fig:gammah} Left panel: fragmentation function ($\xi=\approx -\ln(z)$)
)of the jet tagged by direct $\gamma$ in PHENIX pseudo-rapidity acceptance in p+p and Au+Au collisions.
The solid and dashed line indicate the scaled TASSO data and the scaled MLLA calculation, respectively. Right panel: the $I_{\rm AA}$ calculated with p+p data (filled circles) and scaled TASSO data (open circles).}
\end{flushright}
\end{minipage}
\end{tabular}
\end{figure}

To leading order in pQCD, the energy of a direct $\gamma$ is a good
approximation of the away-side jet energy. Thus one can measure the
in-medium jet fragmentation via $\gamma$-h correlation. Results for
p+p and 0-20\% Au+Au~\cite{gammah} are shown in
Fig.~\ref{fig:gammah}, plotted as a function of fragmentation
variable $\xi=-\ln\left(p_T^a/p_T^{\gamma}
\cos\Delta\phi\right)\approx -\ln(z)$. Small $\xi$ corresponds to
high $p_T$ and vice versa. The fragmentation function from TASSO
(mostly quark jets) and in-medium modified jet fragmentation
function from a MLLA calculation, both for 7 GeV jets, are shown as
lines to compare with the p+p and Au+Au data, respectively. Note
that the TASSO data and MLLA calculation are scaled down by factor
of 10 to match the our data. This factor is needed since the PHENIX
detector has limited $\eta$ acceptance, thus only catches a
fraction of the fragments of the away-side jet.

The $I_{\rm AA}$, or ratio of fragmentation functions of Au+Au to
p+p, is shown as filled circles in the right panel of
Fig.~\ref{fig:gammah}. The $\xi$ range is cut off at 2 due to
limited $p_T$ of the p+p reference data. However, we can extend the
$\xi$ range by calculating $I_{\rm AA}$ using the scaled TASSO data
as reference instead (open circles). The two $I_{\rm AA}$s are
consistent with each other. The data clearly show a suppression at
small $\xi$ (large associated hadron $p_T$) due to jet quenching,
but an enhancement at large $\xi$ (small associated hadron $p_T$)
possibly due to medium response to the quenched jet.

\begin{figure}[h]
\begin{tabular}{lr}
\begin{minipage}{0.7\linewidth}
\begin{flushleft}
\centerline{\includegraphics[width=0.45\textwidth]{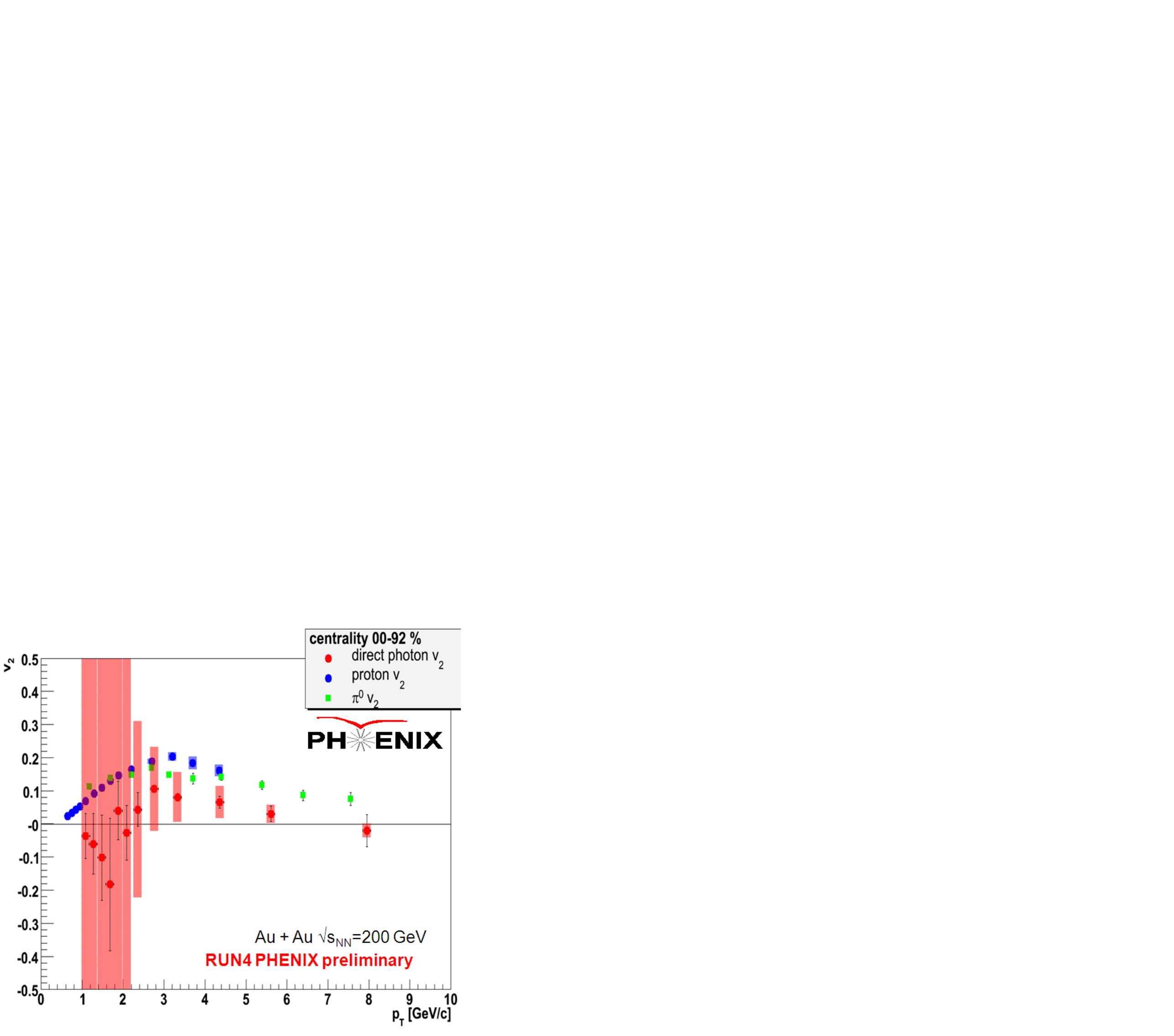}\includegraphics[width=0.6\textwidth]{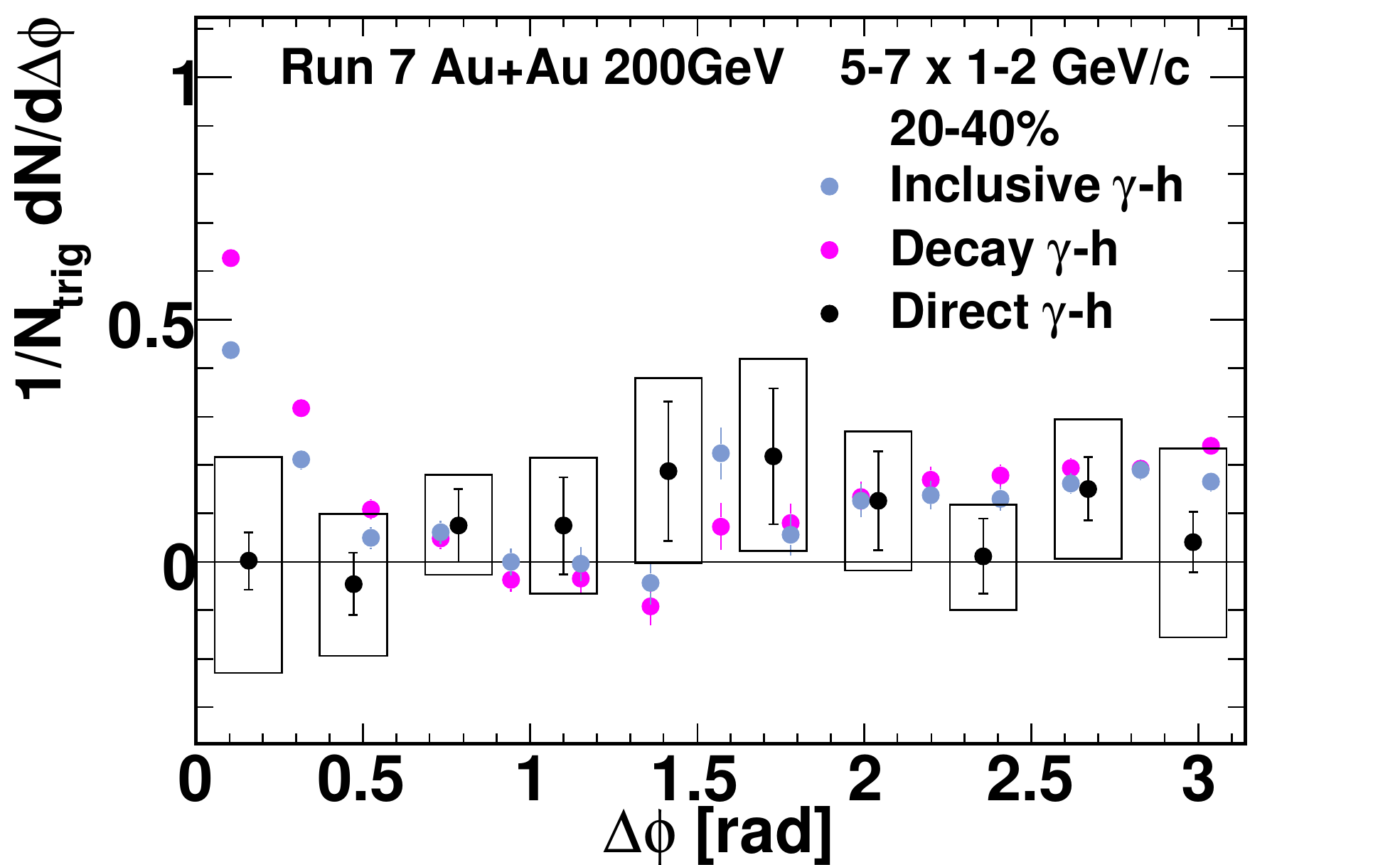}}
\end{flushleft}
\end{minipage}
\hspace*{-0.1in}\begin{minipage}{0.3\linewidth}
\begin{flushright}
\caption{\label{fig:gammah2} Left panel: the direct photon $v_2$; right panel: direct
$\gamma$-hadron azimuthal distribution compared with inclusive and decay $\gamma$-hadron distributions.}
\end{flushright}
\end{minipage}
\end{tabular}
\end{figure}

The enhancement of the associated hadron yield at low $p_T$ was
observed in di-hadron correlations~\cite{Adare:2008cqb}, as a
double hump structure centered around one radian from $\pi$.
However, the interpretation of this enhancement in terms of jet
in-medium response, e.g. Mach cone, is complicated by possible
contributions of E-by-E collective flow
fluctuations~\cite{Takahashi:2009na}. It has been argued that such
fluctuations lead to significant non-zero $v_1$ and $v_3$
components which can mimic such double hump
structure~\cite{Alver:2010gr}. Since the direct $\gamma$ does not
interact strongly with medium, it should have very small flow
signal as indicated by PHENIX preliminary measurements on direct
$\gamma$ $v_2$ (Left panel of Fig.~\ref{fig:gammah2}). Hence the
$\gamma$-h correlation should be relatively unaffected by the $v_n$
contribution, and the associated hadron at low $p_T$ is a robust
measure for the medium response~\cite{Ma:2010dv}. Current
$\gamma$-h correlation indicates some broadening at the away-side
(Right panel of Fig.~\ref{fig:gammah2}), however the statistical
and systematic uncertainties are still too large for a definite
conclusion. With the VTX installed in the next run, PHENIX expects
to have a factor of 20 increase in the effective pair acceptance
for associated hadrons around $\Delta\phi=\pi/2$, a region that is
most crucial for medium response studies.

\section{Modification of fully reconstructed jets}
PHENIX has carried out full jet reconstruction in p+p and Cu+Cu
collisions using the Gaussian filter method~\cite{Lai:2009ai}. This
method is infra-red and collinear safe, and is suitable for limited
acceptance detector. One of the primary challenges for full jet
reconstruction in the heavy ion environment is how to handle the
large underlying background fluctuations, especially at low $p_T$.
PHENIX employed a fake rejection method~\cite{Lai:2009ai}, in which
a certain criterion is defined to suppress fake jets from
background fluctuations that tend to have high multiplicity and
large width. This method can directly suppress fake jets with high
purity, so the systematic error due to underlying event is smaller
than that for the direct background subtraction method. However,
the jet sample passing the rejection criteria are subject to some
efficiency loss and bias, which need to be evaluated carefully.

\begin{figure}[ht]
\centerline{\includegraphics[width=0.9\textwidth]{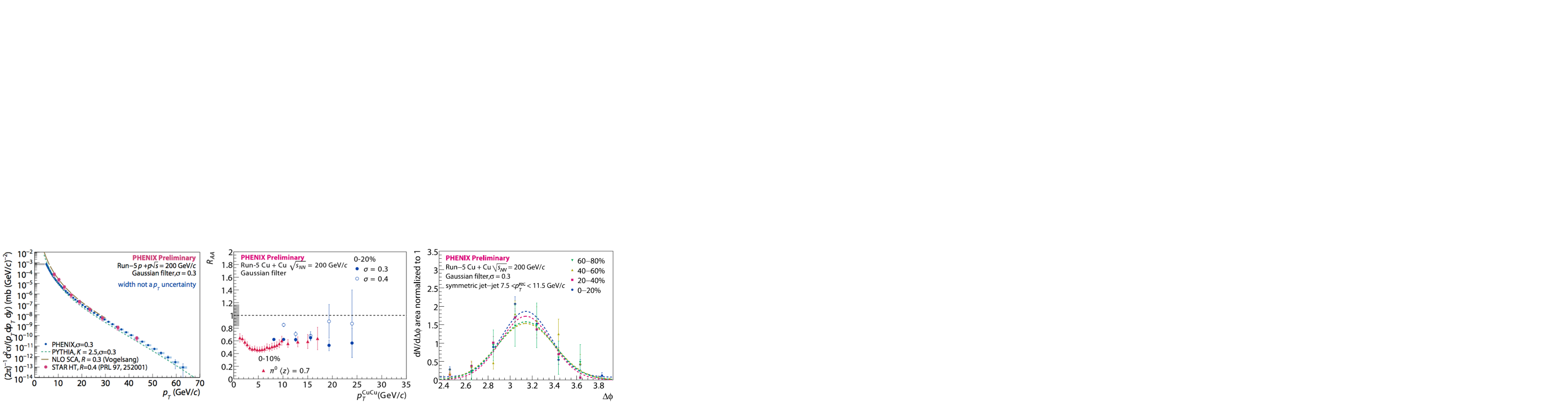}}
\caption[d]
{\label{fig:jet} Left panel: jet spectra in p+p collisions. Middle panel: jet $R_{\rm AA}$ in central Cu+Cu collisions for two jet cones.
Right panel: di-jet azimuthal correlation for several centrality selections in Cu+Cu.}
\end{figure}

Figure~\ref{fig:jet} summarizes the current status of the jet
reconstruction in PHENIX~\cite{Lai:2009ai}. The
left panel shows the full jet spectrum in p+p collisions, unfolded
to particle level, up to 60 GeV. The result is consistent with an
NLO calculation and PYTHIA. The middle panel shows the jet $R_{\rm
AA}$ in central Cu+Cu collisions, unfolded to p+p jet energy scale,
for two different jet cone sizes. Comparison between different jet
cone sizes was argued to directly probe the jet shape
modifications~\cite{Vitev:2009rd}. We see that the $R_{\rm AA}$ for
larger cone size is less suppressed, but the uncertainty also is
much larger presumably due to increased background fluctuation in a
larger cone. The right panel shows the di-jet acoplanarity for
several centrality classes. The widths extracted via Gaussian fits
are consistent across all centrality bins, suggests a small $k_T$
broadening for surviving partons traversing the medium.

\section{Future of jet quenching physics in PHENIX}
The primary goal for jet quenching physics is to obtain a coherent
picture of the interaction of the jets with sQGP. The challenge for
the field is that we are not yet able to simultaneously understand
multiple jet quenching observables, such as $R_{\rm AA}$, $I_{\rm
AA}$, $v_2$, $v_2^{I_{\rm AA}}$, and heavy flavor suppression. In
order to meet this challenge, we need not only more precise
measurements for existing experimental observables, but also the
capability to measure new observables that can provide much more
detailed picture about jet medium interactions. Examples of latter
category include reconstructed jets and di-jets in a broad
acceptance and kinematic range, direct $\gamma$-jet correlation,
and heavy meson tagged jet, just to name a few. PHENIX has planned
aggressive mid-term (2010-2015) and long-term (beyond 2015)
detectors upgrades to fulfill these requirements.

In the mid-term, PHENIX will see the completion of VTX and FVTX
detectors. These detectors should allow us to tag D and B meson
directly, and provide extended acceptance for low $p_T$ charged
hadrons for light/heavy hadron-hadron correlation measurements. The
upgraded data acquisition (Super-DAQ) will take full advantage of
the increased RHIC luminosity. In the long-term, PHENIX plans to
replace the existing outer central detectors with a compact large
acceptance EMCal and HCal, which together with VTX, FVTX,
additional tracking layers and high DAQ rate, will allow us to
measure jets, dijets, heavy flavor jets, and direct photon-jet
correlations in a broad kinematic ranges. We refer more detailed
discussion to~\cite{jacak}.
\section{Summary}
PHENIX has made several new measurements on single hadron and
di-hadron correlation observables. By combining information from
multiple observables, we are now able to better constrain jet
quenching mechanisms and discriminate different models. We show
that the $R_{\rm AA}$ measurement, with increased precision,
suggests a gradual increase at high $p_T$. We find that the $v_2$
at high $p_T$ exceeds the pQCD predictions, suggesting a
non-trivial path length dependence of the energy loss. We also find
that the $I_{\rm AA}$ is less suppressed than $R_{\rm AA}$, $I_{\rm
AA}>R_{\rm AA}$, yet its anisotropy is larger than single hadrons,
$v_2^{I_{\rm AA}}>v_2$. This result is rather non-trival given the
anti-correlation between $R_{\rm AA}$ and $v_2$, and between
$I_{\rm AA}$ and $v_2^{I_{\rm AA}}$.

PHENIX also made good progresses on $\gamma$-hadron correlations
and full jet reconstruction. These measurements are challenging
either due to their low rate ($\gamma$-hadron) or large
underlying-event background fluctuation (full jet reconstruction).
By extending the $\gamma$-hadron correlations to low associated
$p_T$, we observed strong evidence of enhancement due to energy
dissipation of quenched jets. We have measured jet spectra in p+p
and Cu+Cu collisions, and explored the modification of jet shape
and di-jet broadening. These measurements will benefit tremendously
from future detector and luminosity upgrade of the PHENIX.
\section*{Acknowledgement}
This work is supported by NSF under award number PHY-1019387.





\bibliographystyle{elsarticle-num}



\end{document}